\documentclass{revtex4}

\usepackage{amsmath}
\usepackage{graphicx}

\begin{document}

\title{Relations Involving Static Quadrupole Moments of $2^+$ states and B(E2)'s}
\author{S. Yeager}
\affiliation{Department of Physics and Astronomy, Rutgers University
Piscataway, New Jersey 08854}
\author{L. Zamick}
\affiliation{Department of Physics and Astronomy, Rutgers University
Piscataway, New Jersey 08854}

\begin{abstract}
We define the ``quadrupole ratio'' $r_Q = \dfrac{Q_0(S)}{Q_0(B)}$ where $Q_0(S)$ is the intrinsic quadrupole moment obtained from the static quadrupole moment of the $2_1^+$ state of an even-even nucleus and $Q_0(B)$ the intrinsic quadrupole moment obtained from $B(E2)_{0 \rightarrow 2}$. In both cases we assume a simple rotational formula connecting the rotating frame to the laboratory frame. The quantity $r_Q$ would be one if the rotational model were perfect and the energy ratio $E(4)/E(2)$ would be $10/3$. In the simple vibrational model, $r_Q$ would be zero and $E(4)/E(2)$ would be two. There are some regions where the rotational limit is almost met and fewer where the vibrational limit is also almost met.  For most cases, however, it is between these two limits, i.e. $0 < |r_Q| < 1$. There are a few cases where $r_Q$ is bigger than one, especially for light nuclei.
\end{abstract}

\maketitle

\section{Introduction}

In previous works \cite{robinson06,escuderos} a certain ratio was defined involving the static quadrupole moment of a $2^+$ state in an even-even nucleus and the $B(E2)_{0 \rightarrow 2}$.  It was defined in such a way so that for a perfect rotor, the ratio was one and for a simple vibrator, the ratio was zero.

In the rotational model the $Q(2^+)$ and $B(E2)_{0 \rightarrow 2}$ are described by one parameter, the intrinsic quadrupole moment Q$_0$. We will however define two different operational static quadrupole moments.

\begin{subequations}
\begin{gather}
B(E2)_{I1 \rightarrow I2} = \frac{5}{16\pi} Q_0^2(B) | \langle I_1 \;\; K \;\; 20 \;\; \mid \;\; I_2 \;\; K \rangle |^2  \\ Q(I) = \frac{3K^2 - I(I+1)}{(I+1)(2I+3)} Q_0(S)
\end{gather}
\end{subequations}

Where

\begin{subequations}
\begin{gather}
Q_0(B) = \sqrt{B(E2)} \sqrt{\frac{16 \pi}{5}} \\
Q_0(S) = - \frac{7}{2}Q(I)
\end{gather}
\end{subequations}

For $K=0$, $I_1=0$, $I_2=2$ we obtain

\begin{subequations}
\begin{gather}
B(E2)_{0 \rightarrow 2} = \frac{5}{16\pi}Q_0^2(B) \\
Q(2^+) = -\frac{2}{7}Q_0(S)
\end{gather}
\end{subequations}

The quadrupole ratio is

\begin{align}
r_Q = \frac{Q_0(S)}{Q_0(B)} &= -\frac{7}{2}\sqrt{\frac{5}{16\pi}}\frac{Q(2^+)}{\sqrt{B(E2)_{0 \rightarrow 2}}} \\
&= -1.1038\frac{Q(2^+)}{\sqrt{B(E2)_{0 \rightarrow 2}}} \nonumber
\end{align}

Thus, we have expressed the quadrupole ratio in terms of quantities measured in the laboratory.  For $Q(2^+)$ we use the reference of Stone \cite{stone05} and for B(E2), the tables of Raman et al. \cite{raman01}.

The main difference from the previous works is that we now consider all nuclei for which $Q(2^+)$ has been measured. In previous works, only light nuclei were considered.  Looking at all nuclei, as we do here, will give us a new perspective.

\section{Choices made in getting $Q(2^+)$}

It is much easier to measure $B(E2)_{0 \rightarrow 2}$ than it is to measure $Q(2^+)$.  The values of $B(E2)$ in Raman's table form a concensus made by Raman when more than one measurement was made on a nucleus.

For $Q(2^+)$, Stone does not present an evaluated result, and wisely so, because there is a wide variation in some cases.  What was done in this work was to select the latest measurement of a given $Q(2^+)$. Of course it is not always true that the latest measurement is the best one.  Hence, if later, more definate measurements are made, we will have to alter the figures.  We feel that the large scale impression of the results will still stand.

Note that Stone's table does not contain measured values of $Q(2_1^+)$ beyond $^{206}$Pb. Hence in this analysis we cannot comment on nuclei beyond doubly magic $^{208}$Pb such as the actinide nuclei and the transuranic nuclei.

\section{Results}

In Table I we present for even-even nuclei the magnitude ofthe quadrupole ratio and $E(4)/E(2)$, the ratio of energies. Again, in the pure rotation model $r_Q$ is equal to 1 and $E(4)/E(2) = \dfrac{4 \times 5}{2 \times 3} = 3.33$. In a simple vibrational limit the quadrupole moment would be zero (i.e. the static quadrupole moment would vanish) and $E(4)/E(2)=2$.

\includegraphics[width=17.5cm]{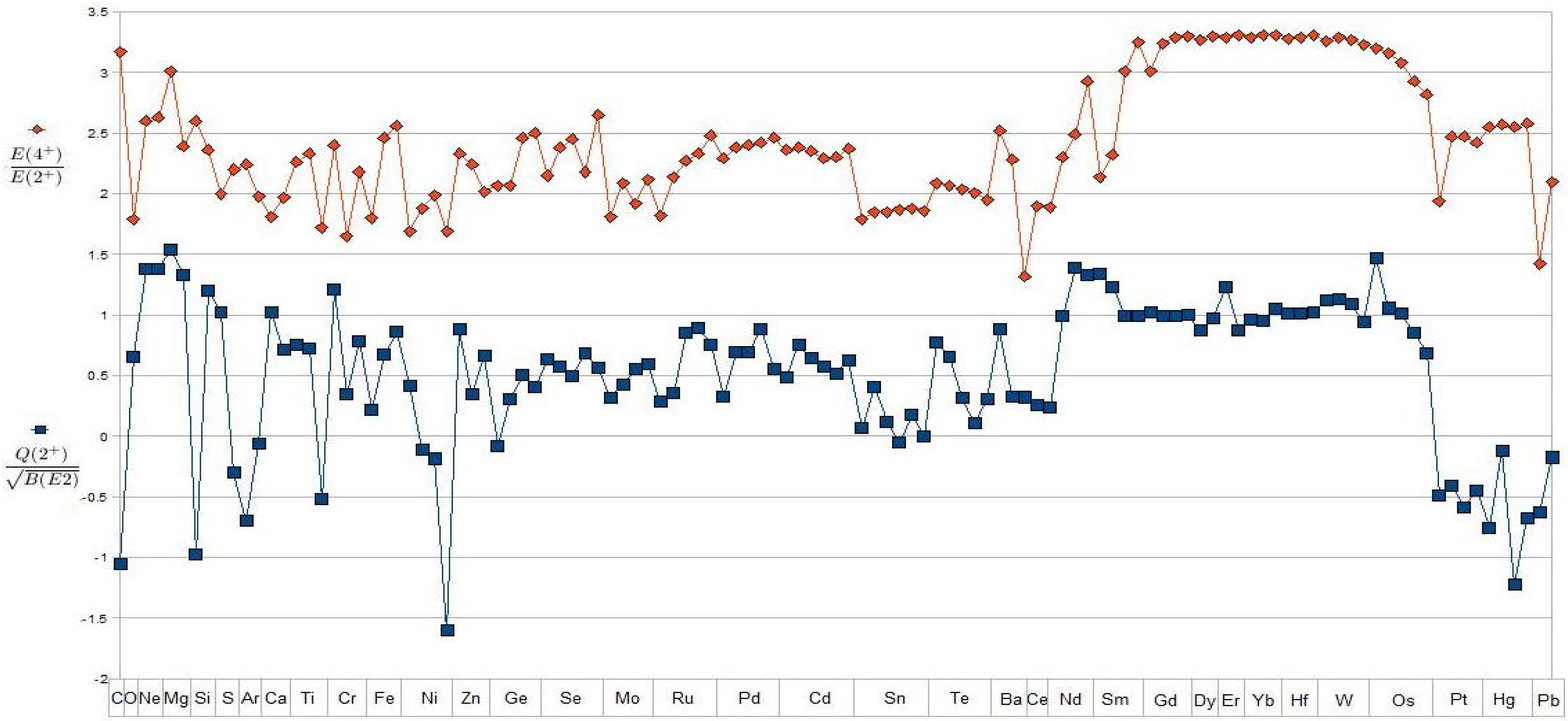}

The results are also shown in Fig 1.  The upper curve is the ratio $E(4)/E(2)$ and the lower curve is the ratio involving $Q(2^+)$ and $B(E2)$.

For light nuclei there are several examples where the quadrupole ratio is even larger than one.  There is a midrange region from about 70 to 110 where the quadrupole ratio is around 0.5 and where $E(4)/E(2)$ is between two and three.  This is a region which is about halfway between vibrational and rotational.  Then we hit an island 112 and 120 where thhe simple vibrational model is almost realized. The nuclei in this band include $^{112, 118, 120, 122, 124}$Sn and $^{128}$Te. The ratio here for $Q(2^+)$ is almost zero and $E(4)/E(2)$ is close to two.

As we move to heavier nuclei, the rotational model is realized in a band from about 152 to 186, with the quadrupole ratio close to one and $E(4)/E(2)$ close to 10/3. These nuclei include $^{146, 148, 150}$Nd, $^{152, 154}$Sn, Gd isotopes, $^{160}$Dy, $^{164}$Dy, Er isotopes, Yb isotopes, $^{182, 184, and 186}$W, most Os isotopes, and $^{202}$Hg.

We can also consider other models. For example we can use the $f_{7/2}$ configuration for $^{42}$Ca. This single particle model is the antithesis of the rotational model. Yet it yields a value $r_Q = 0.78$. This compares with the experimental value of 1.023.

\section{Closing Remarks}

This work represents an expansion of the 2006 work of Robinson et al. \cite{robinson06}. We here consider essentially all nuclei where $Q(2^+)$ has been measured. Since that work, there have been other works of relevance e.g. that of Bertsch et al. \cite{bertsch07} in which the Gogny interaction was used to calculate $Q(2^+)$ and $B(E2)$ as well as work by Sabbey et al. \cite{sabbey07}. Early works using Skyrme H.F. were performed by Jaqaman et al. \cite{jaqaman84} who also considered hexadecapole models. Recent work by Sarriguren et al. \cite{sarriguren} should be noted as well as the phase transitions in the platinum isotopes by Morales et al.\cite{morales08}. They consider the heavier nuclei -- isotopes of Yb, Hf, W, Os, and Pt -- where there are many changes from prolate to oblate. Also mentioned in the 2006 work, Zelevinsky and Volya \cite{zelevinsky04} noted that with random matrices they obtained with a high probability that the quadrupole ratio $r_Q$ (as we define it) is either one or zero. Why this is so is not clear.

We thank the Aresty program at Rutgers for their support. We thank Yitzhak Sharon for his interest, Noemie Koller, Jolie Cizewski, and Gulhan Gurdal for critical but cogent comments, and Gerfried Kumbartzki for his help.

\bibliographystyle{unsrt}
\bibliography{Sources}

\begin{thebibliography}{10}

\bibitem{robinson06}
S.~J. Q. Robinson{,} A. Escuderos{,} L. Zamick{,}~P. von Neumann{-}Cosel{,} A.
  Richter{,} R. W.~Fearick.
\newblock {\em Phys. Rev.}, C73:037306, 2006.

\bibitem{escuderos}
A.~Escuderos{,}~L. Zamick.
\newblock {\em LANL Preprint{,} arXiv{:} nucl{-}th}, page 10605030.

\bibitem{stone05}
N.~J. Stone.
\newblock {\em {Atomic Data and Nuclear Data Tables}}, 90:75, 2005.

\bibitem{raman01}
S.~Raman{,} C. W. Nestor Jr.{,}~P. Tikkaner.
\newblock {\em Atomic Data and Nuclear Data Tables}, 78:1, 2001.

\bibitem{bertsch07}
G.~F. Bertsch{,} M. Girod{,} O. Kenn{,} S. Hilaire{,} J. P. Delaroche{,}~H.
  Goutte{,} and S.~Peru.
\newblock {\em Phys Rev Lett}, 99:032502, 2007.

\bibitem{sabbey07}
B.~Sabbey{,} M. Bender{,} G.~F. Bertsch{,} and P-H. Heenan.
\newblock {\em Phys Rev}, C75:044305, 2007.

\bibitem{jaqaman84}
H.~R. Jaqaman and L.~Zamick.
\newblock {\em Phys Rev}, C30:1719, 1984.

\bibitem{sarriguren}
P.~Sarriguren{,}~R. Rodriguez-Guzman{,} and L.~M. Robledo.
\newblock {\em LANL preprint nucl-th}, page 0806.2079.

\bibitem{morales08}
I.~O. Morales{,} A. Frank{,} C.~E. Varma{,} and P.~VanIsacher.
\newblock {\em Phys Lett B}, 551:249, 2003.

\bibitem{zelevinsky04}
V.~Zelevinsky and A.~Volya.
\newblock {\em Phys Rev}, 391, 2004.

\end{thebibliography}
\end{document}